# Specific features of band structure and optical anisotropy of $Cu_2CdGeSe_4$ quaternary compounds

M.G. Brik[a,1], O.V. Parasyuk[b], G.L. Myronchuk[c], I.V. Kityk[d]

[a] Institute of Physics, University of Tartu, Riia 142, Tartu 51014, Estonia
[b] Department of Chemistry, Eastern European National University, Voli 13, Lutsk 43025, Ukraine
[c] Department of Physics, Eastern European National University, Voli 13, Lutsk 43025, Ukraine
[d] Institute of Materials Science and Engineering, Technical University of Czestochowa, Al. Armii Krajowej 19, 42-200 Czestochowa, Poland

**Abstract**

The complex theoretical and experimental studies of the band structure and the optical functions of the $Cu_2CdGeSe_4$ quaternary crystals are reported. The benchmark band structure calculations were performed using the first-principles methods. As a result, the structural, electronic, optical and elastic properties of $Cu_2CdGeSe_4$ were calculated in the general gradient approximation (GGA) and local density approximation (LDA). The calculated dielectric function and optical absorption spectra exhibit some anisotropic behavior. Detailed analysis of the band energy dispersion and the effective space charge density helped in establishing the origin of the band structure anisotropy. All calculated properties are compared with the experimental data. An additional comparison with a similar crystal of $Cu_2CdGeSe_4$ allowed to reveal the role played by the anions (S or Se) in formation of the optical properties of these two materials.

## 1. Introduction

The quaternary $A_2^I B^{II} C^{IV} X_4$ phases, where A is Cu, $B^{II}$ is Zn, Cd, Hg; $C^{IV}$ is Si, Ge, Sn; X is S, Se, Te, belong to normal-valence diamond-like semiconductors. The majority of these materials crystallize in the tetragonal and orthorhombic structures, which are derivatives of the sphalerite and wurtzite structures, respectively [1, 2, 3]. Non-central symmetry of these compounds' lattices is one of the reasons why these materials are widely used in nonlinear optics [4]. This fact is probably also the reason for their piezo- and pyro-electric properties [5]. The results of investigation of the phase equilibria in the $Cu_2GeSe_3$–CdSe quasi-binary system [6] show that the quaternary compound melts incongruently at 1103 K and has, in addition, a polymorphous transition at 878 K [7] or 875 K [8].

---

[1] Corresponding author. E-mail: brik@fi.tartu.ee Phone: +372 7374751

According to Ref. [9], $Cu_2CdGeSe_4$ single crystals were obtained using directional solidification, some of their physical properties were investigated and reported. $Cu_2CdGeSe_4$ is a typical p-type semiconductor with optical energy gap value equal to 1.29 eV [10] or 1.20 eV [11] and strong optical anisotropy. Its microhardness is equal to 1.90 GPa [9] or 2.32 GPa [12], and its density is equal to 5.45 g/cm$^3$ [12]. The results of the crystal structure investigation of two polymorphous modifications of the $Cu_2CdGeSe_4$ compound are given in Ref. [13].

Recently, we performed detailed experimental and first-principles studies of a similar compound $Cu_2CdGeS_4$ [14]. In the present work we report on the results of the experimental (crystal growth and spectroscopy) and first-principles (optimized crystal structure, electronic, optical and elastic properties) studies of $Cu_2CdGeSe_4$ in comparison with a similar compound of $Cu_2CdGeS_4$ with an aim of clarification of the role played by the anions in formation of the physical properties of these materials.

In the next section we describe the procedure of the samples preparation and spectroscopic measurements, then we proceed with description of the calculation settings and results of the calculations of the structural, electronic and optical properties at the ambient pressure. A separate section deals with the pressure effects and their influence on the structural and optical properties of $Cu_2CdGeSe_4$. Finally, the paper is concluded with a short summary.

## 2. Samples preparation and details of spectroscopic measurements

$Cu_2CdGeSe_4$ crystals were grown by a horizontal variation of Bridgman method. The starting composition for the growth was selected from the field of the primary crystallization of the quaternary phase (38 mole percent of CdSe and 62 mole percent of $Cu_2GeSe_3$) according to the investigated $Cu_2GeSe_3$–CdSe phase diagram [8]. The calculated amounts of elements of at least 99.999 wt.% purity with the total weight of 15 g were loaded into a quartz container with conical bottom. The ampoule was evacuated and sealed. The crystal growth was performed in two steps. First, the container was placed in a shaft-type furnace and heated at the rate of 30 K/h to 1200 K. The batch was kept at this temperature for 6 h with periodic vibration mixing and then cooled to the room temperature over 24 h. Second, the container with the alloy was placed in a horizontal furnace inclined at an angle of ~10º. The container was then pulled horizontally at a rate of 1 mm/h. The growth zone and the annealing zone temperatures were 1200 K and 770 K, respectively. The temperature gradient at the crystallization interface did not exceed 1.7 K/mm. After reaching the isothermal zone at 770 K, the crystals were annealed for 250 hours and then cooled down to room temperature at a rate of 100 K/day. The obtained boule consisted of two parts, a $Cu_2CdGeSe_4+Cu_2GeSe_3$ eutectic and a crystalline portion which contained several single-crystalline blocks of $Cu_2CdGeSe_4$, the largest of which was 8×4×3 mm$^3$ in size.

## 3. Crystal structure and details of calculations

Obtained $Cu_2CdGeSe_4$ crystals were characterized by XRD spectra recorded by the DRON 4-13 diffractometer using $CuK_\alpha$ radiation. It was determined that they crystallize in the stannine structure. The diffraction pattern clearly indicates the tetragonal structure in agreement with the results given in Refs. [4-9]. All computations were performed using the CSD software package [15]. The refinement of coordinates and isotropic temperature displacement parameters using the full-profile Rietveld method yielded a good value of the fit factor $R_I$=0.0582. Experimental and theoretical (obtained by using the calculated parameters) diffraction patterns of the $Cu_2CdGeSe_4$ compound as well as the differential pattern are shown in Fig. 1.

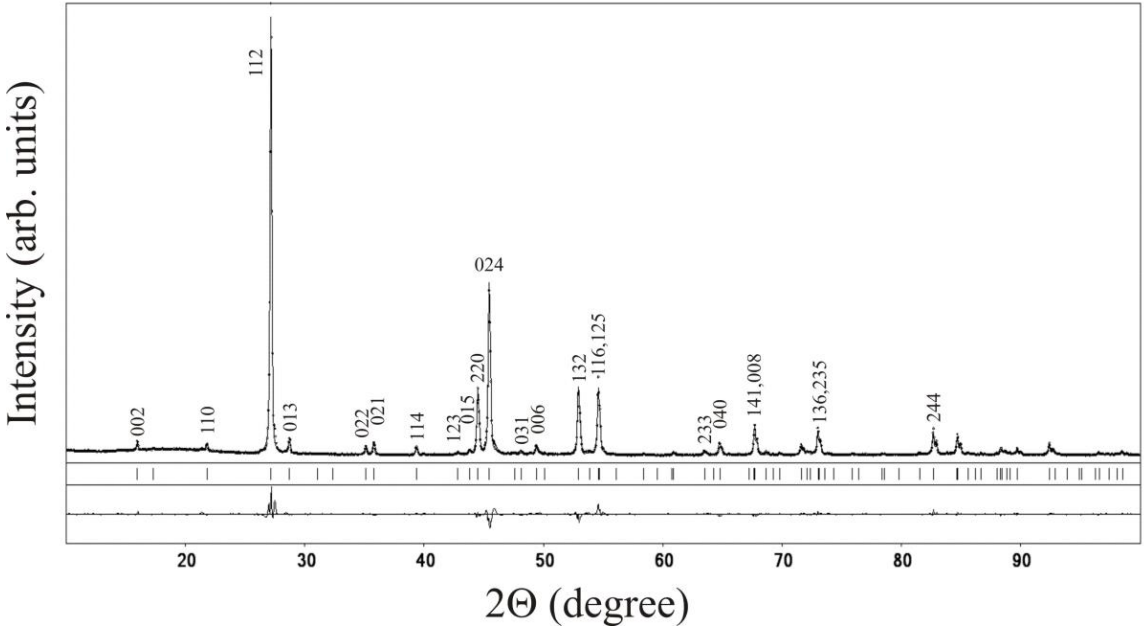

Fig. 1. Experimental and calculated X-ray diffraction patterns and their difference (the first line above the horizontal axis) for the studied $Cu_2CdGeSe_4$ single crystal.

The CASTEP module [16] of Materials Studio was used to calculate the structural, electronic, optical and elastic properties of $Cu_2CdGeSe_4$. The initial structural data for this compound were taken from Ref. [17]. According to this reference, there are two structural modifications of $Cu_2CdGeSe_4$. At low temperature, it crystallizes in the I-42m space group (No. 121) with two formula units in one unit cell. At high temperature (over 875 K [8]) the structure is changed to that one described by the $Pmn2_1$ space group. The low-temperature (LT) phase was studied in the present paper; its unit cell is depicted in Fig. 2, whereas Fig. 3 shows the Brillouin zone (BZ) for the primitive cell of this structure. The generalized gradient approximation (GGE) with the Perdew-Burke-Ernzerhof [18] and the local density approximation (LDA) with the Ceperley-Alder-Perdew-Zunger (CA-PZ) functional [19, 20] were used to treat the exchange-correlation effects. The Monkhorst-Pack k-points grid was chosen as 8×8×4. The cut-off energy, which determines the size of the plane-wave basis set, was 320 eV. The convergence criteria were as follows: $5\times10^{-6}$ eV/atom for energy, 0.01 eV/Å for maximal force, 0.02 GPa for maximal stress and $5\times10^{-4}$ Å for maximal displacement. The electronic configurations were the following: $3d^{10}4s^1$ for Cu, $4d^{10}5s^2$ for Cd, $4s^24p^2$ for Ge, and $4s^24p^4$ for Se, and the ultrasoft pseudopotentials were used for all chemical elements. The calculations were performed for a primitive cell of $Cu_2CdGeSe_4$.

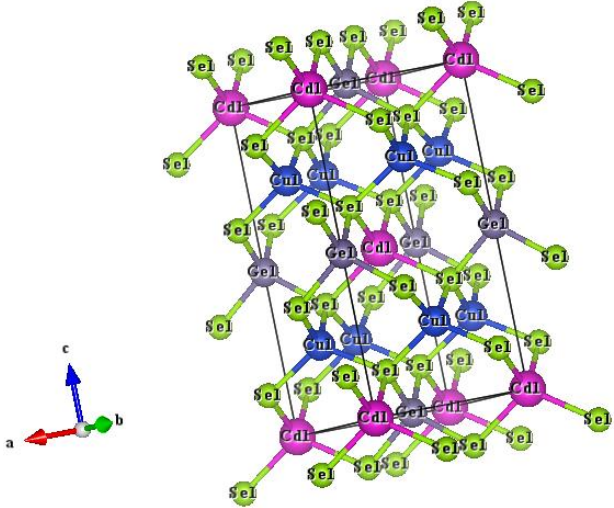

Fig. 2. One unit cell of $Cu_2CdGeSe_4$. Drawn by VESTA [21].

## 4. Results of calculations: structural, electronic and optical properties

Table 1 contains a summary of the experimental and optimized structural data for $Cu_2CdGeSe_4$. Comparison of the calculated and experimental results yields good agreement between them (the maximal relative difference between the experimental and optimized lattice parameters is 0.9 % in the case of the GGA calculations, and 1.9 % in the case of the LDA calculations), which makes a firm basis for reliability of the subsequent analysis of the electronic, optical and elastic properties of this material.

Table 1. Summary of the structural data for the LT phase of $Cu_2CdGeSe_4$.

| | Exp. [17] | | | Calculations | | |
|---|---|---|---|---|---|---|
| $a$, Å | 5.7482(2) | | | 5.8008[a], 5.6462[b], 5.75216[c], 5.5000[d] | | |
| $c$, Å | 11.0533(3) | | | 11.0845[a], 10.8429[b], 10.9454[c], 11.0605[d] | | |
| | Atomic coordinates | | | | | |
| | x/a | y/b | z/c | x/a | y/b | z/c |
| Cu | 0.5 | 0 | 0.25 | 0.5 | 0 | 0.25 |
| Cd | 0 | 0 | 0 | 0 | 0 | 0 |
| Ge | 0 | 0 | 0.5 | 0 | 0 | 0.5 |
| Se | 0.2705(2) | 0.2705(2) | 0.1337(1) | 0.26904[a], 0.26890[b], 0.26143[c] | 0.26904[a], 0.26890[b], 0.26143[c] | 0.13732[a], 0.13716[b], 0.13911[c] |

[a] This work, GGA [b] This work, LDA [c] Ref. [22] [d] Ref. [23]

According to Refs. [24, 25], the experimental band gap of $Cu_2CdGeSe_4$ is 1.29 eV at 300 K. It is considerably lower than the energy gap of $Cu_2CdGeS_4$ (2.05 eV, [14, 26]). The calculated band gap values were 0.020 eV (GGA) and 0.032 eV (LDA), which are substantially underestimated with respect to the experimental result. Such a band gap underestimation is a common feature of the DFT-based methods. In order to overcome it, a scissor operator can be used. The latter simply shifts upward the conduction band to make the band gap match the corresponding experimental data. In our case, the value of such a shift was equal to 1.27 eV (GGA) and 1.26 eV (LDA).

Fig. 4 presents both calculated band structures with taking into account the scissor operator. The band gap is of a direct nature, since the maximum of the valence band and the minimum of the conduction band are both situated at the BZ center, similarly to the case of $Cu_2CdGeS_4$ [14]. Comparing the band structures of $Cu_2CdGeS_4$ and $Cu_2CdGeSe_4$, one can notice considerably higher degree of the band energy dispersion (or curvature of the calculated energy bands) in the latter case, in particularly along the G-X-P line of the BZ, which is in accordance to the general crystallochemistry presenting in the Fig. 2.

This may reflect a higher iconicity of the Ge-Se bonds with respect to the analogous bonds Ge-S for $Cu_2CdGeS_4$ [14] ones. Moreover, one can conclude that for both crystals the maximal effective masses and lower motilities of the charge carriers are achieved near the BZ center. This is a crucial factor for the understanding of the carrier motilities for this type of quaternary semiconductor crystals.

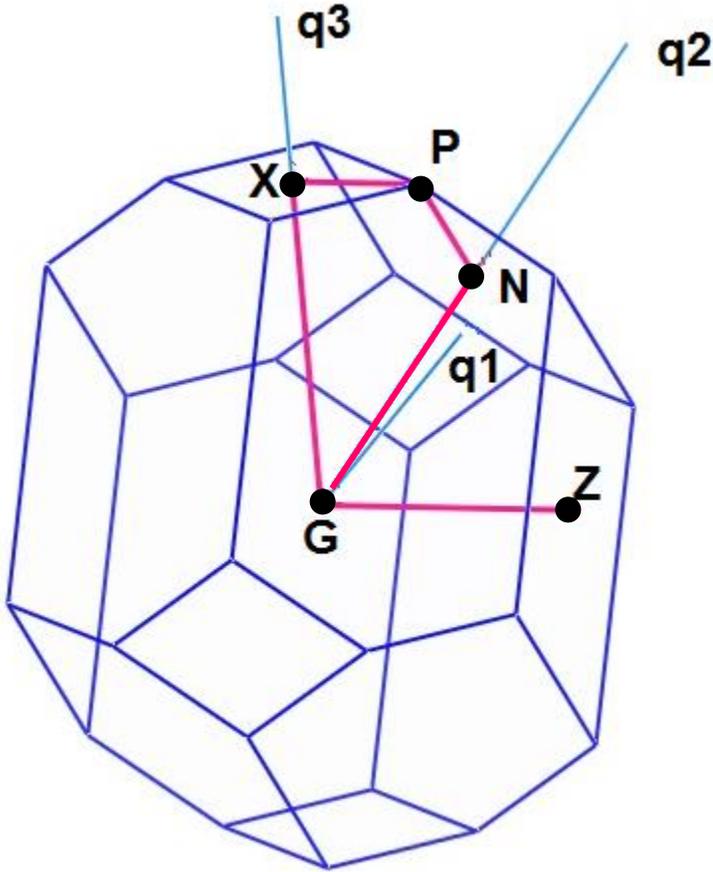

Fig. 3. The Brillouin zone for a primitive cell of $Cu_2CdGeSe_4$ single crystal. The coordinates of the special points of the Brillouin zone are (in units of the reciprocal lattice vectors); Z (1/2, 1/2, -1/2); G (0, 0, 0); X (0, 0, 1/2), P(1/4, 1/4, 1/4), N(0, 1/2, 0).

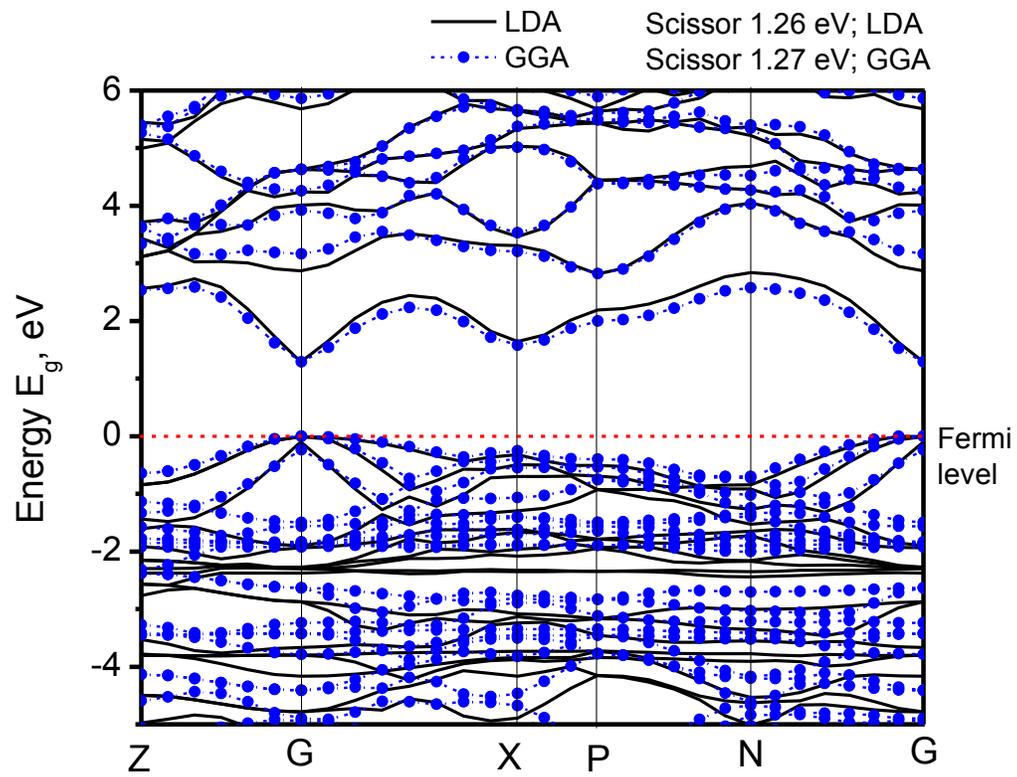

Fig. 4. Calculated band structure of $Cu_2CdGeSe_4$. The value of the scissor operator is indicated.

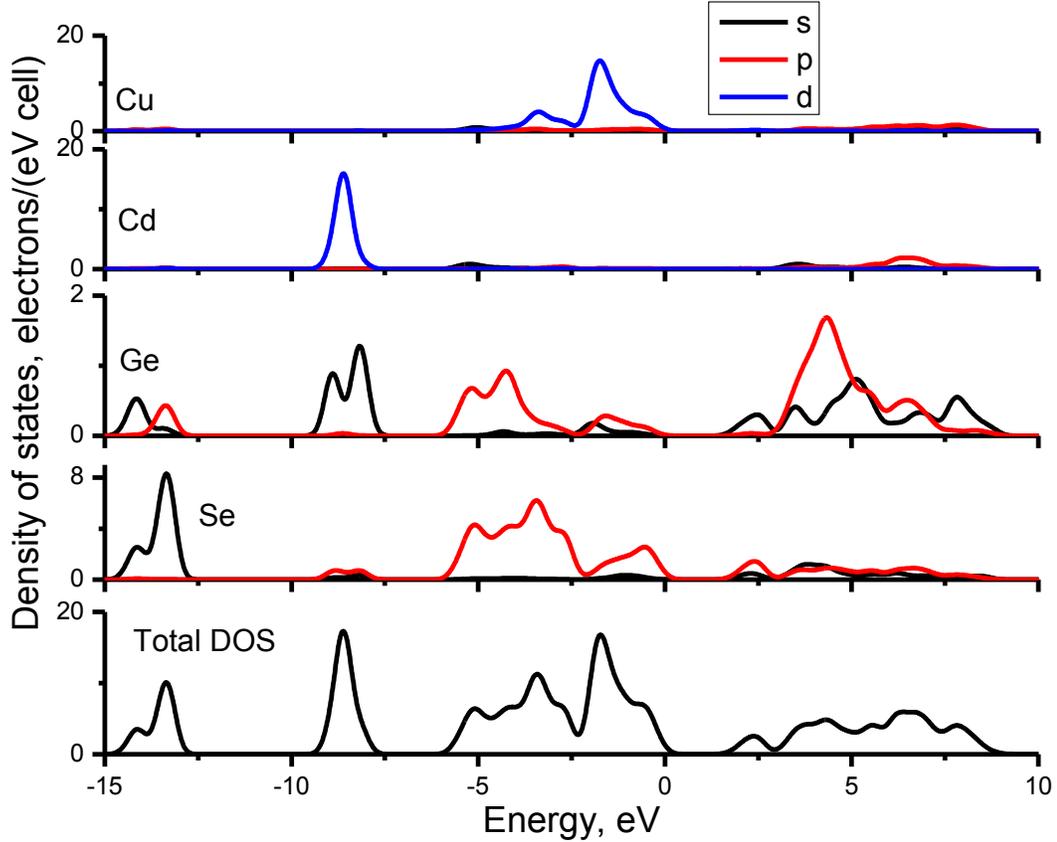

Fig. 5. Calculated density of states (DOS) diagrams for $Cu_2CdGeSe_4$.

Fig. 5 shows the partial density of states diagrams, which allow to find the composition of the calculated electronic bands as follows: the conduction band is very wide, with the width of about 7 eV. It is made basically of the Ge 4s, 4p states, with admixture of the Se 4s, 4p states. Their shapes are more delocalized than for the case of the $Cu_2CdGeS_4$ crystals [14] due to different degree of ionicities for $Cu_2CdGeSe_4$ with respect to $Cu_2CdGeS_4$. The valence band, whose width is about 6 eV, has a complicated structure and is formed by the Cu 3d, Se 4p states, with a slight admixture of the Ge 4s, 4p states due to hybridization and the two principal valence sub-bands are almost overlapped contrary to the $Cu_2CdGeS_4$ crystals, where a distinct gap between -2 eV and -3 eV in the valence band was found to exist [14]. The Cu 3d states exhibit two clearly pronounced localized peaks at about -3.4 eV and -1.7 eV, which can be interpreted as the splitting of the 3d states in a tetrahedral crystal field into the $e_g$ and $t_{2g}$ states, respectively. A similar situation was encountered in $Cu_2CdGeS_4$ crystals [14]. The Cd 4d states produce a sharp peak at about -8.6 eV, and the Ge 4s states are situated at -8.2 eV and -9 eV. Finally, the lowest energetic bands are formed by the Se 4s states with the peaks at about -14 eV and -13 eV.

The optical properties of a solid can be analyzed if its dielectric function is known. The imaginary part Im($\varepsilon(\omega)$) of a dielectric function $\varepsilon(\omega)$ (directly related to the absorption spectrum of a solid) is calculated by numerical integrations of matrix elements of the electric dipole operator between the occupied states $\Psi_k^c$ in the valence band and empty states $\Psi_k^v$ in the conduction band:

$$\text{Im}(\varepsilon(\omega)) = \frac{2e^2\pi}{\omega\varepsilon_0} \sum_{k,v,c} \left|\left\langle \Psi_k^c \left| \vec{u}\cdot\vec{r} \right| \Psi_k^c \right\rangle\right|^2 \delta(E_k^c - E_k^v - E), \qquad (1)$$

where $\vec{u}$ is the polarization vector of the incident electric field, $\vec{r}$ and $e$ are the electron's position vector and electric charge, respectively, $E = \hbar\omega$ is the incident photon's energy, and $\varepsilon_0$ is the vacuum dielectric permittivity. The summation in Eq. (1) is carried out over all states from the occupied and empty bands, with their wave functions obtained in a numerical form after optimization of the crystal structure.

The real part Re($\varepsilon(\omega)$) of the dielectric function $\varepsilon$, which determines the dispersion properties and refractive index values, is estimated in the next step by using the Kramers-Kronig relation:

$$\text{Re}(\varepsilon(\omega)) = 1 + \frac{2}{\pi} \int_0^\infty \frac{\text{Im}(\varepsilon(\omega'))\omega' d\omega'}{\omega'^2 - \omega^2} \qquad (2)$$

The calculated real and imaginary parts of the dielectric function for $Cu_2CdGeSe_4$ are shown in Fig. 6. Taking the square root from the values of $\text{Re}(\varepsilon(\omega))$ in the limit of zeroth energy, one can estimate the value of the refractive index, which is then 3.08 (LDA) and 3.10 (GGA). Additionally one can see a red shift of the plasmon resonances, which occur when $\text{Re}(\varepsilon(\omega))$=0, with respect to the sulfur containing crystals. This property reflects a different affinity of the particular elements – S and Se.

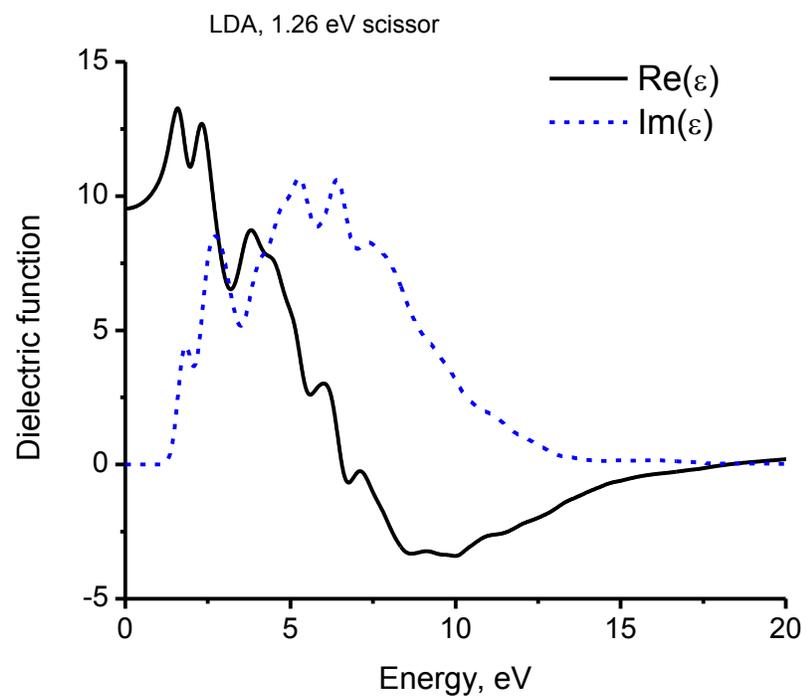

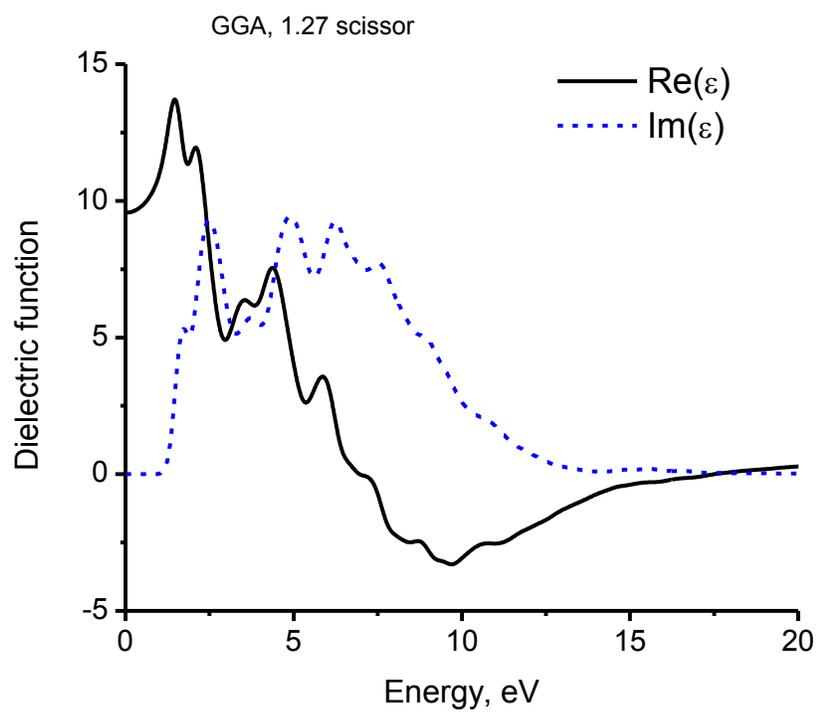

Fig. 6. Calculated dielectric function for $Cu_2CdGeSe_4$.

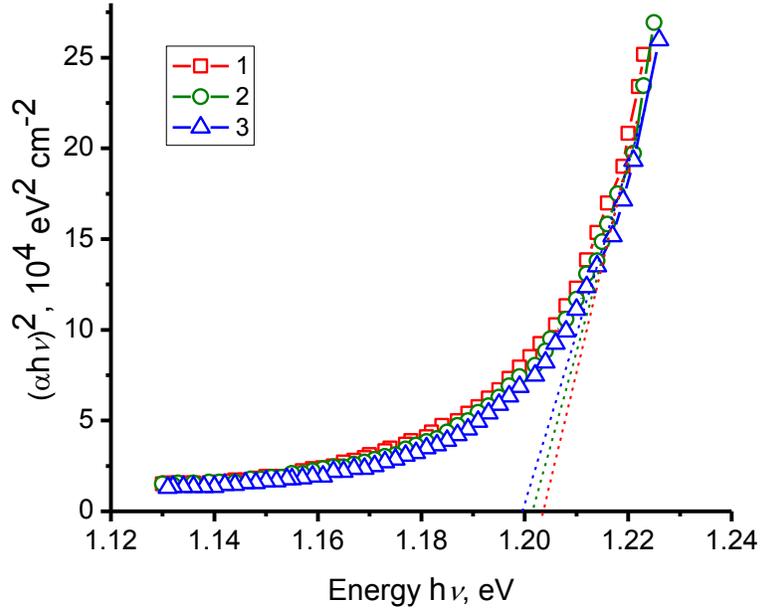

Fig. 7. Absorption edge for the $Cu_2CdGeSe_4$ single crystal for different polarizations: 1 – non-polarized light; 2 – $\bar{E} \parallel c$; 3 – $\bar{E} \perp c$.

Fig. 7 shows the experimental polarized absorption spectra of the $Cu_2CdGeSe_4$ crystal near the fundamental absorption edge.

The absorption coefficient $\alpha$ was calculated using the following equation [27]:

$$\alpha = \frac{1}{d} \ln\left\{ \frac{(1-R)^2}{2T} + \sqrt{\left[\frac{(1-R)^2}{2T}\right]^2 + R^2} \right\} \tag{3}$$

where $d$ is the sample thickness; $T = I/I_0$ is the transmission coefficient; $R$ is the reflection coefficient ($R = 0.35$–$0.40$ depending on the crystal composition).`

To estimate the band energy gap for the direct ($E_{gd}$) allowed transitions, the $(\alpha h\nu)^2 = f(h\nu)$ lines in Fig. 7 were extrapolated to the value $(\alpha h\nu)^2 = 0$; then the value of 1.20 eV of the $Cu_2CdGeSe_4$ band gap at 300 K was obtained. The degree of anisotropy between two absorption directions determined by the differences in the energy gaps (in the plane of the crystals, i.e. $E \parallel c$ and $E \perp c$) is about 0.4 eV which is a bit higher (0.28–0.34 eV) with respect to the $Cu_2CdGeS_4$ [14].

Generally, the inter-band transitions for different light polarizations have some anisotropy of the transition dipole moment's magnitudes, which are closely related to the corresponding charge density distribution around ions in a crystal lattice. In the case of the investigated crystal, this is caused by the space charge distribution of the chemical bonds which form the principal valence bands.

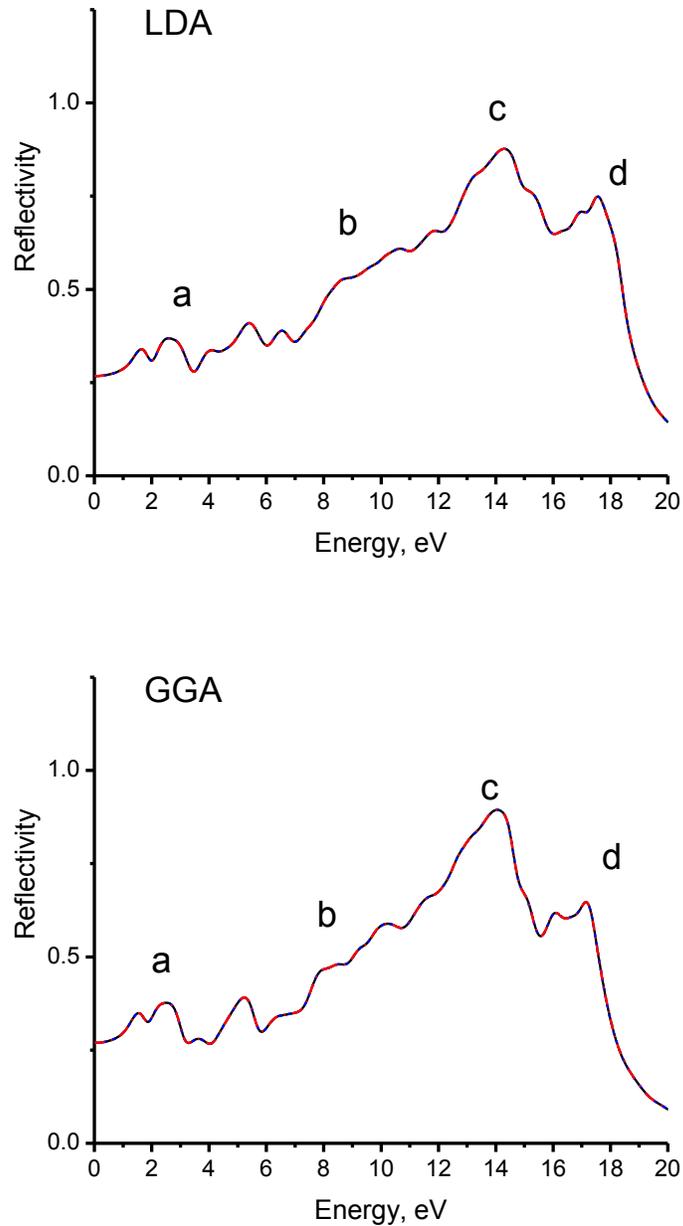

Fig. 8. Calculated polarized reflectivity spectra of $Cu_2CdGeSe_4$.

Fig. 8 shows the calculated reflectivity spectra, which turned out to be practically independent of the polarization of light.

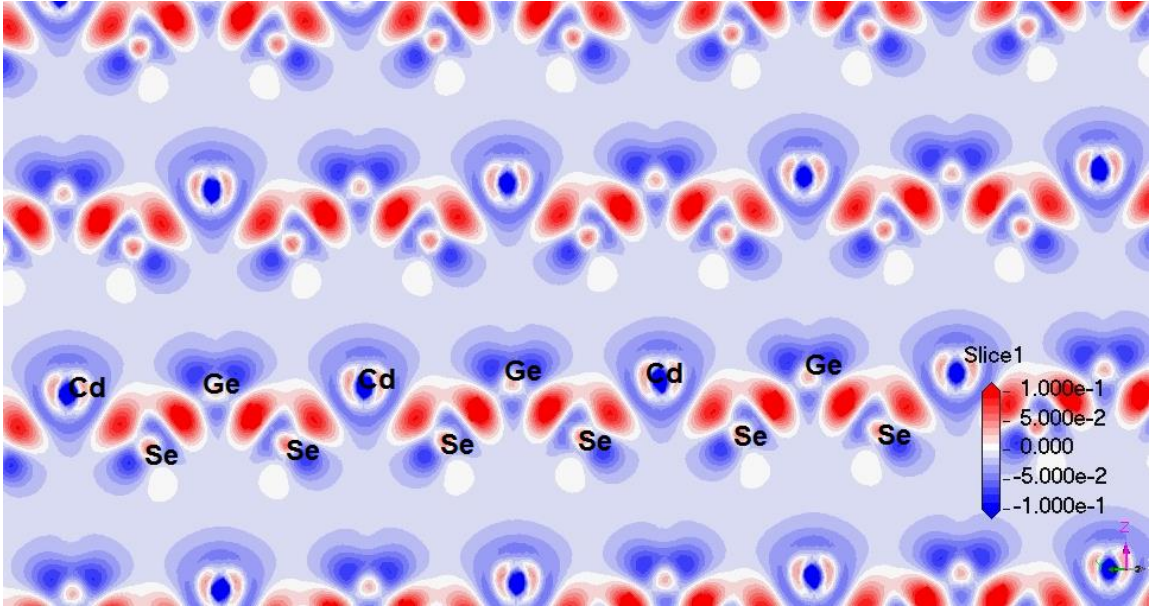

Fig. 9. Calculated cross-section of the electron density difference in the plane containg the Ge, Cd, and Se ions for $Cu_2CdGeSe_4$.

The calculated cross-sections of the electron density difference in the space between the atoms in the $Cu_2CdGeSe_4$ crystal lattice is depicted in Fig. 9. One can clearly see formation of the chemical bonds between the Ge–Se and Cd–Se atoms, since the regions in the space between these atoms gain electron density. The effective Mulliken populations of all ions are given in Table 2; the effective charges of all ions are very different from those formally expected from the chemical formula. It is obvious that the charge density distribution around the Se atoms is substantially more delocalized if compared to the charge distribution around S atoms in $Cu_2CdGeS_4$ [14].

Table 2. Effective Mulliken populations and charges (GGA/LDA) for $Cu_2CdGeSe_4$.

|    | s | p | d | Total | Charge |
|----|-----|-----|-----|-------|--------|
| Cu | 0.68/0.70 | 0.84/0.94 | 9.79/9.79 | 11.31/11.43 | -0.31/-0.43 |
| Cd | 0.84/0.92 | 0.93/1.01 | 9.98/9.97 | 11.75/11.90 | 0.25/0.10 |
| Ge | 1.17/1.10 | 2.13/2.25 | 0.00/0.00 | 3.30/3.35 | 0.70/0.65 |
| Se | 1.67/1.63 | 4.41/4.35 | 0.00/0.00 | 6.08/5.98 | -0.08/0.02 |

## 5. Results of calculations: pressure effects on the structural, electronic and optical properties

It is already well-known that it is possible considerably change the properties of a solid by applying to it a hydrostatic pressure. The pressure effects on the structural and electronic properties of $Cu_2CdGeSe_4$ were modeled by performing the crystal structure optimization and calculating the electronic and optical properties at the elevated hydrostatic pressure, in the range from 0 to 20 GPa with a step of 5 GPa. The calculated structural parameters and band gaps (with the scissor operators) are given in Table 3.

Table 3. Calculated band gaps (in eV, with scissor operators) and lattice parameters $a$, $c$ (all in Å) for $Cu_2CdGeSe_4$ at different pressures.

| Pressure, GPa | GGA | | | LDA | | |
|---|---|---|---|---|---|---|
| | Band gap | $a$ | $c$ | Band gap | $a$ | $c$ |
| 0  | 1.29  | 5.800796 | 11.084461 | 1.292 | 5.646217 | 10.842924 |
| 5  | 1.305 | 5.676961 | 10.772368 | 1.313 | 5.550994 | 10.585954 |
| 10 | 1.404 | 5.597887 | 10.497308 | 1.458 | 5.483469 | 10.357964 |
| 15 | 1.549 | 5.532389 | 10.280391 | 1.565 | 5.436892 | 10.135204 |
| 20 | 1.656 | 5.491823 | 10.040136 | 1.612 | 5.417869 | 9.863425  |

The data from Table 3 are visualized in Figs. 10–12. Since the band gap of the studied crystal is a direct one, it is increasing with pressure. Its dependence on pressure is well described by a linear function with the slope of about 0.020 eV/GPa in both GGA and LDA approximations.

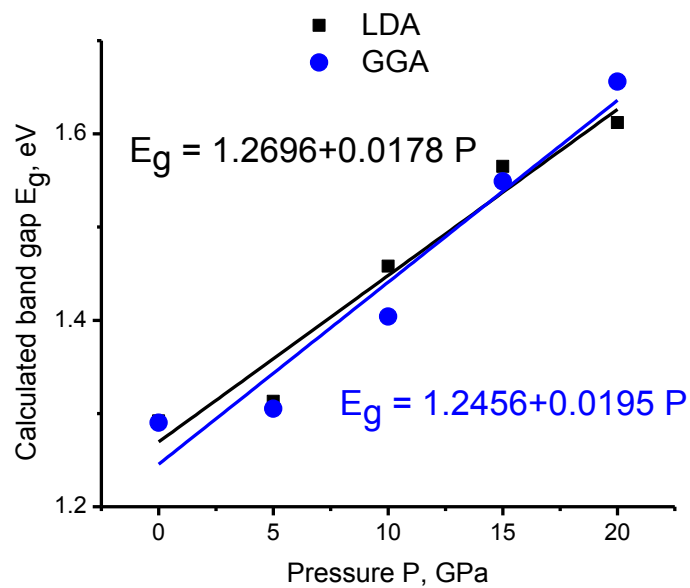

Fig. 10. Calculated band gap (symbols) and a linear fit (solid lines) as function of pressure $P$ for $Cu_2CdGeSe_4$.

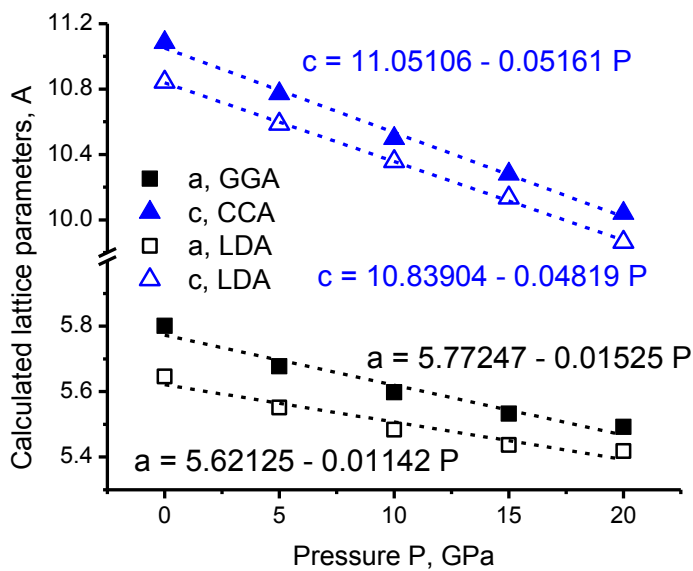

Fig. 11. Calculated lattice parameters (symbols) and their linear fits (dashed lines) as function of pressure $P$ for $Cu_2CdGeSe_4$.

Dependencies of the calculated lattice parameters on pressure are shown in Fig. 11. All $a$ and $c$ parameters decrease linearly with pressure, with the largest (in absolute value) pressure coefficient for the $c$ parameter in both GGA and LDA calculations, which means that the largest compressibility of $Cu_2CdGeSe_4$ is realized along a direction with the largest value of the lattice parameter (along the $c$ crystallographic axis in this case). This is similar to $Cu_2CdGeS_4$, however the line slopes are different: about 0.02 Å/GPa $Cu_2CdGeS_4$ [14] vs 0.05 Å/GPa ($Cu_2CdGeSe_4$, this work).

Finally, after the lattice parameters at various values of hydrostatic pressure are found, the volume $V$ of a unit cell can be readily calculated to be fitted to the Murnaghan equation of state [28]:

$$\frac{V}{V_0} = \left(1 + \frac{PB'}{B}\right)^{-1/B'}, \qquad (4)$$

where $V_0$ denotes the unit cell volume at the ambient pressure, $B$ is the bulk modulus, and $B' = dB/dP$ is its pressure derivative.

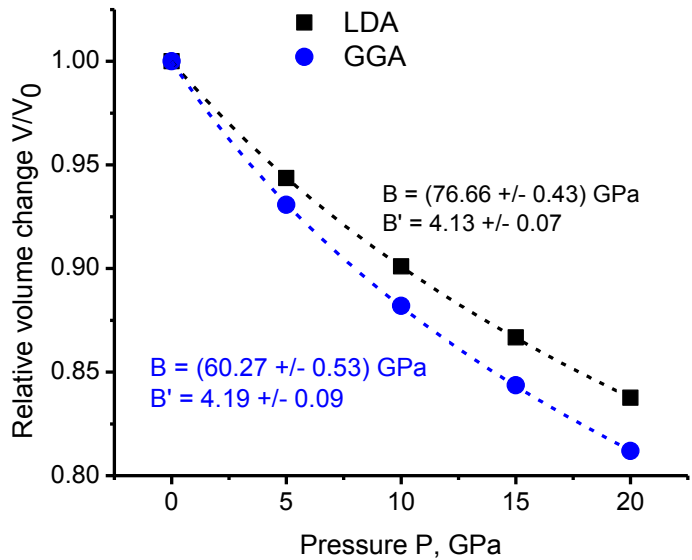

Fig. 12. Calculated dependence of the relative volume change $V/V_0$ (symbols) and fits to the Murnaghan equation of state (dashed lines) as function of pressure $P$ for $Cu_2CdGeSe_4$.

The results of application of Eq. (4) to the calculated variation of the $Cu_2CdGeSe_4$ unit cell volume are shown in Fig. 12. Estimations of the bulk modulus give the values of about 77 GPa (LDA) and 60 GPa (GGA), which are somewhat lower with respect to 91 GPa (LDA) and 69 GPa (GGA) for $Cu_2CdGeS_4$ [14]. To the best of our knowledge, these are the first estimates of the bulk modulus for $Cu_2CdGeSe_4$.

## 6. Conclusions

Experimental and theoretical studies of a quaternary semiconductor $Cu_2CdGeSe_4$ are reported in the present paper. Details of the crystal growth, XRD structure investigations and optical spectra measurements are all described. In particular, the grown $Cu_2CdGeSe_4$ single crystals were obtained using directional solidification and the horizontal modified Bridgman method.

For the first time, the electronic band structure, optical features and related parameters of $Cu_2CdGeSe_4$ were studied using the first-principles methods. Comparison of the calculated and experimental lattice parameters yields good agreement (the maximal relative difference between the experimental and optimized lattice parameters is 0.9 % in the case of the GGA calculations, and 1.9 % in the case of the LDA calculations), which confirms reliability of the subsequent analysis of the electronic, optical and elastic properties of this material. The calculated band gap values were equal to 0.020 eV (GGA) and 0.032 eV (LDA); considerable underestimation of these band gaps was overcome by using the scissor factor (1.27 eV (GGA) and 1.26 eV (LDA)).

The band gap of $Cu_2CdGeSe_4$ is of a direct nature; the electronic states are more delocalized than in the case of $Cu_2CdGeS_4$ crystals. The valence band, whose width is about 6 eV, does not exhibit any sub-bands dual structure, like in $Cu_2CdGeS_4$. From the calculated dielectric functions, we found a slight red-shift of the plasmon resonances in $Cu_2CdGeSe_4$ with respect to $Cu_2CdGeS_4$. In a certain way, it can be also related to the decreased band gap value of $Cu_2CdGeSe_4$ and its higher covalency.

We also report the first estimation of the bulk modulus for the title compound as 77 GPa (LDA) and 60 GPa (GGA), which show that the replacement of S by Se (i.e. going from $Cu_2CdGeS_4$ to $Cu_2CdGeSe_4$) makes a compound be somewhat softer due to decrease of the bulk modulus. It was also evidenced by the calculated pressure coefficients of the lattice parameters: they decrease faster with increasing pressure in $Cu_2CdGeSe_4$.

The results obtained in this paper clearly illustrate how the properties of quaternary semiconductors can be modified by changing the anions; this can be important and interesting for effective engineering of materials with desired properties.

## Acknowledgments

MGB appreciates the financial support from (i) the European Social Fund's Doctoral Studies and Internationalisation Programme DoRa, (ii) the European Union through the European Regional Development Fund (Center of Excellence 'Mesosystems: Theory and Applications', TK114) and (iii) the Marie Curie Initial Training Network LUMINET, grant agreement no. 316906. Dr. G.A. Kumar (University of Texas at San Antonio) is thanked for allowing us to use the Materials Studio package.